\documentclass[aps,showpacs,twocolumn]{revtex4}
\usepackage[dvips]{graphics,graphicx}
\begin{document}
\title{Dynamical instability, Chaos, and Bloch oscillations of Bose-Einstein
condensates in tilted optical lattices}
\author{Andrey R. Kolovsky}
\affiliation{Max-Planck-Institut f\"ur Physik komplexer Systeme,
             D-01187 Dresden}
\affiliation{Kirensky Institute of Physics, Ru-660036 Krasnoyarsk.}
\date{\today}

\begin{abstract}
We study the Bloch dynamics of a Bose-Einstein condensate of cold
atoms by using the formalism of the discrete nonlinear Schr\"odinger
equation. Depending on the static force magnitudes the system is shown to exhibit
two qualitatively different regimes of Bloch oscillations -- exponential decay 
for the static force magnitude less than some critical value (defined by
the condensate density) and quasiperiodic oscillations in the opposite case.
The relation of these regimes to the onset of chaos in the system is discussed.
\end{abstract}
\maketitle

\section{Introduction}
Recently much attention has been payed to Bloch oscillations (BO)
of Bose-Einstein condensate (BEC) of cold atoms in the optical lattices
\cite{Mors01,Cris03,Wu03,Roat04,Zhen04,Trom01,Thom03,prl57,pre60,prl61}.
It was found that, unlike the case of a dilute atomic gas, where
the atom-atom interactions can be neglected, BO of BEC rapidly decay.
A plausible explanation of this phenomenon is suggested by the
dynamical instability of the Gross-Pitaevskii equation in
the presence of periodic potential 
(see papers \cite{Cris03,Wu03} and references therein).
In this paper we revisit this approach, assuming for simplicity
the relatively deep optical lattices, where the Gross-Pitaevskii equation can
be approximated by the discrete nonlinear Schr\"odinger equation (DNLSE).
It is shown below that a strong static force can actually suppress 
the dynamical instability of the DNLSE and, as a consequence,
BOs become not-decaying. We indicate a condition on the
critical magnitude of the static force separating the decaying and
not-decaying regimes of BO which, for a fixed static force, can be also
formulated as a condition on the critical density of BEC. This result
is well supported by the numerical simulation of the system dynamics,
performed for the realistic parameters of the present-days laboratory
experiments with cold atoms.

\section{Dynamical instability}
According to DNLSE the (mean field) Bloch dynamics of a BEC
is described by the system of coupled nonlinear equations
\begin{equation}
\label{1}
i\hbar\dot{a}_l=-\frac{J}{2}(a_{l+1}+a_{l-1})
+g|a_l|^2a_l+dFla_l \;.
\end{equation}
In Eq.~(\ref{1}) $a_l(t)$ is the complex amplitude of a BEC of
atoms localised in the $l$th well of the optical potential,
$J$ is the hopping or tunnelling matrix elements (who's value
is uniquely defined by the lattice depth), $d$ the lattice
period, $F$ magnitude of the static force, and $g$ the nonlinear
parameter given by the product of the microscopic interaction
constant $W$ and the filling factor ${\bar n}$ (mean number
of atoms per lattice site), $g={\bar n}W$. For a nonuniform
density ${\bar n}$ will mean the peak density.

\begin{figure}[t!]
\center
\includegraphics[width=8.5cm, clip]{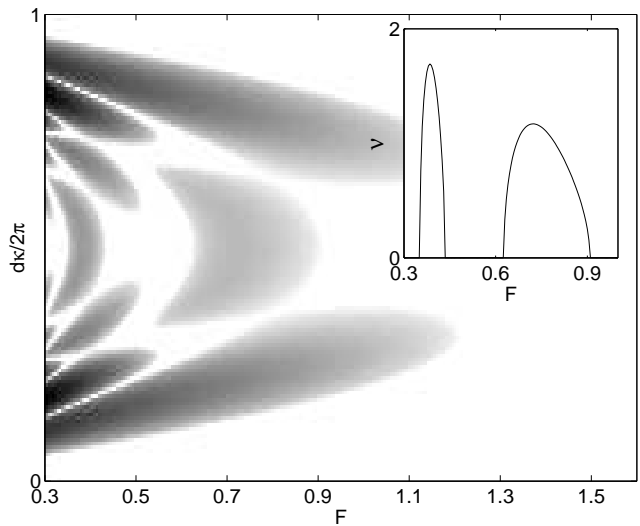}
\caption{Increment of the dynamical (modulation) instability $\nu$ as the 
function of the static force magnitude $F$ and the quasimomentum $\kappa$ for
$g=0.4$. The lattice period $d$ and the tunnelling constant $J$ are set
to unity. The inset shows $\nu=\nu(F)$ for $\kappa=\pi/d$ (edge of the
Brillouin zone).}
\label{fig1}
\end{figure}

Let us first consider the case of uniform initial conditions
$a_l(0)=1$. (Note that we do not normalise $\sum_l |a_l|^2$
to unity.) Subsequently using the gauge and Fourier transformation,
$b_\kappa=L^{-1/2}\sum_{l=1}^L \exp(i\kappa l-i\omega_B l t)a_l$,
Eq.~(\ref{1}) can be written in the form,
\begin{equation}
\label{2}
i\hbar\dot{b}_\kappa=-J\cos(d\kappa-\omega_Bt)b_\kappa
\end{equation}
\begin{displaymath}
+\frac{g}{L}\sum_{\kappa_1,\kappa_2,\kappa_3} b_{\kappa_1}
b^*_{\kappa_2}b_{\kappa_3} \delta(\kappa-\kappa_1+\kappa_2-\kappa_3) \;,
\end{displaymath}
where $\kappa$ is the quasimomentum ($-\pi/d\le\kappa<\pi/d$),
and $\omega_B=dF/\hbar$ Bloch frequency. For the given initial
condition Eq.~(\ref{2}) has trivial solution
\begin{equation}
\label{3}
b_0(t)=\exp\left(i\frac{J}{dF}\sin(\omega_B t)-i\frac{g}{\hbar}t
\right) \sqrt{L} \;,\quad b_{\kappa\ne0}(t)=0 \;,
\end{equation}
which is no other than the celebrated BO,
where the kinetic and Stark energies of the system oscillate
according to the cosine law with the Bloch period $T_B=2\pi/\omega_B=h/dF$.
However, for $g\ne0$ the solution (\ref{3}) can be unstable with respect
to small perturbations. Following the standard approach we linearise
Eq.~(\ref{2}) around the solution (\ref{3}), which leads to the system 
of linear equations,
\begin{eqnarray}
\nonumber
i\hbar{\dot b}_{+\kappa}=-J\cos(d\kappa-\omega_Bt)b_{+\kappa}
+2\frac{g}{L}|b_0|^2 b_{+\kappa}+\frac{g}{L}b_0^2 b^*_{-\kappa} \;, \\
\label{4}
i\hbar{\dot b}_{-\kappa}=-J\cos(d\kappa+\omega_Bt)b_{-\kappa}
+2\frac{g}{L}|b_0|^2 b_{-\kappa}+\frac{g}{L}b_0^2 b^*_{+\kappa} \;,
\end{eqnarray}
where $b_{\pm\kappa}(0)$ are arbitrary small. Substituting here $b_0(t)$
from Eq.~(\ref{3}) and integrating Eq.~(\ref{4}) in time,
we have $b_{\pm\kappa}(t)\sim \exp(\nu t)b_{\pm\kappa}(0)$,
with $\nu$ given by the logarithm of the maximal eigenvalue
of the Floquet matrix
\begin{equation}
\label{5}
U=\widehat{\exp}\left[-i\frac{g}{\hbar}\int_0^{T_B}\left(
\begin{array}{cc}
1&f(t)\\-f^*(t)&-1
\end{array}
\right)dt\right] \;,
\end{equation}
\begin{displaymath}
f(t)=\exp\left[i\frac{2J}{dF}[1-\cos(d\kappa)]\sin(\omega_B t)\right] \;,
\end{displaymath}
where the hat over the exponent denotes the time ordering.
It is easy to see that the matrix $U$ is parametrised by the product
$d\kappa$, the ratio $g/dF$, and the ratio $J/dF$. Thus, without lost of
generality, we can set two parameters ($J$ and $d$ in what follows) to unity.
A typical dependence of the increment $\nu$ on the quasimomentum
$\kappa$ and the static force magnitude $F$ is depicted
in Fig.~\ref{fig1}. It is seen in the figure that there is a critical
value $F_{cr}\approx1.2$ above which $\nu\equiv0$ and, hence, 
the solution (\ref{3}) is stable. Scanning over different $g$, we
found $F_{cr}\approx3g$ for $0.1\le g\le1$ \cite{Zhen04}.

Suppression of the dynamical instability is easy to check numerically.
These numerical simulations, performed for
different lattice size $L$ and periodic boundary conditions,
undoubtly indicate existence of the critical $F$. Moreover, this
is also valid for the non-uniform initial conditions
$|a_l|^2=\exp(-l^2/\sigma^2)$, realized in the experiment. Although in 
this case and for $F>F_{cr}$ BO slowly decay (with subsequent revivals, see
Fig.~\ref{fig6} below), this decay is due to the trivial
dephasing \cite{Trom01} and fundamentally differs from the exponentially
fast destruction of BO at $F<F_{cr}$, caused by the dynamical instability.

\section{Two-site model}
\label{sec3}
\begin{figure}[t!]
\center
\includegraphics[width=4.0cm, clip]{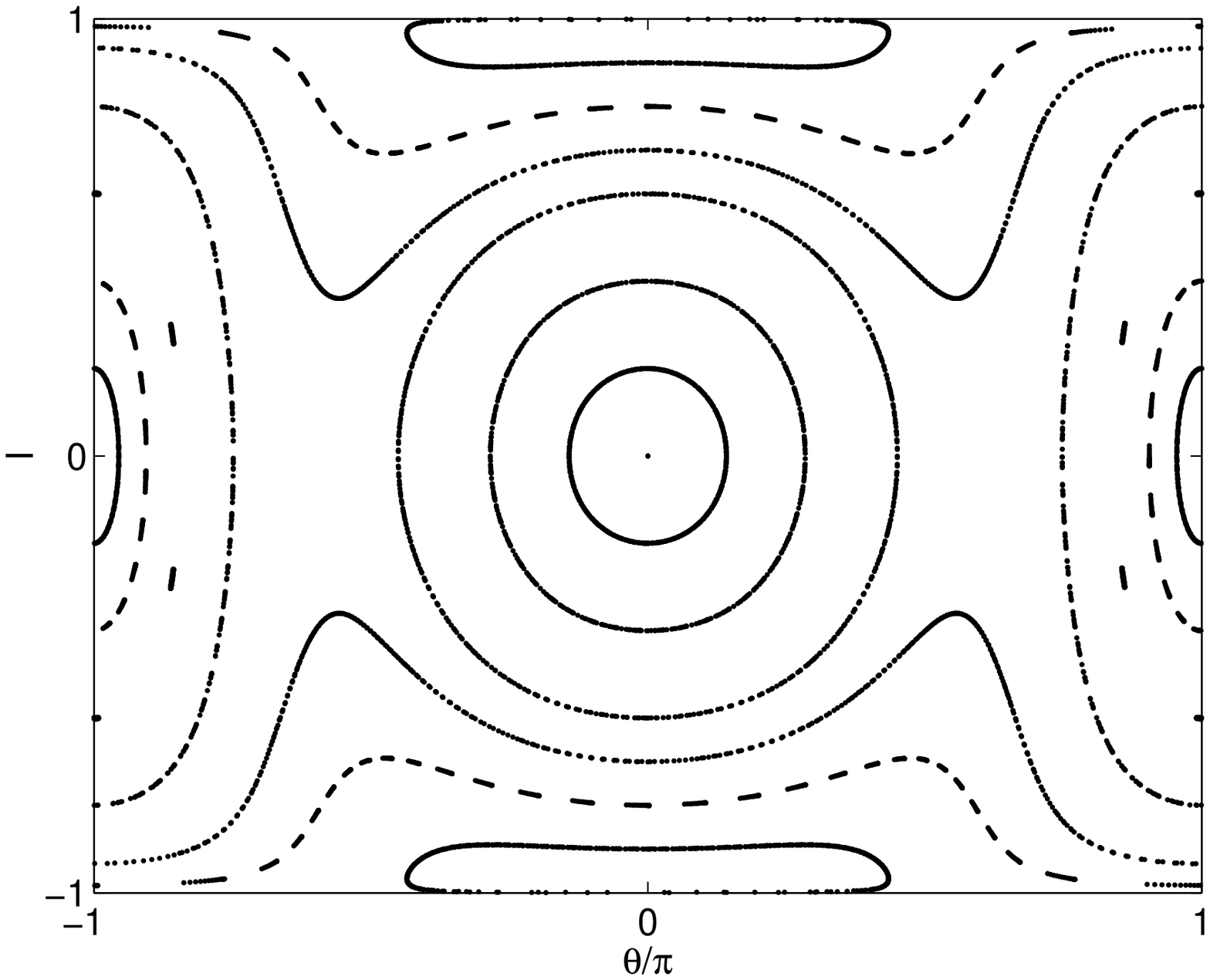}
\includegraphics[width=4.0cm, clip]{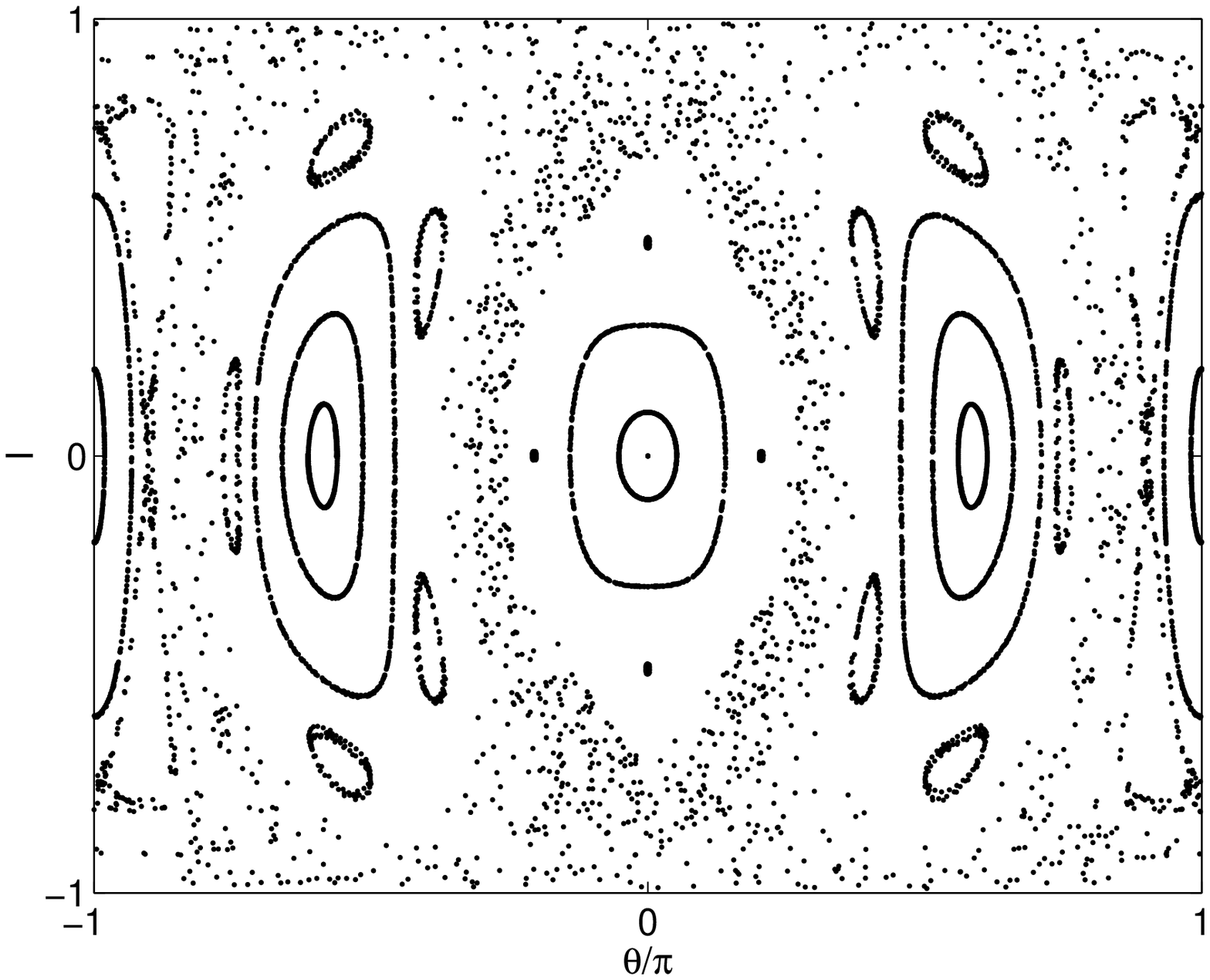}
\includegraphics[width=4.0cm, clip]{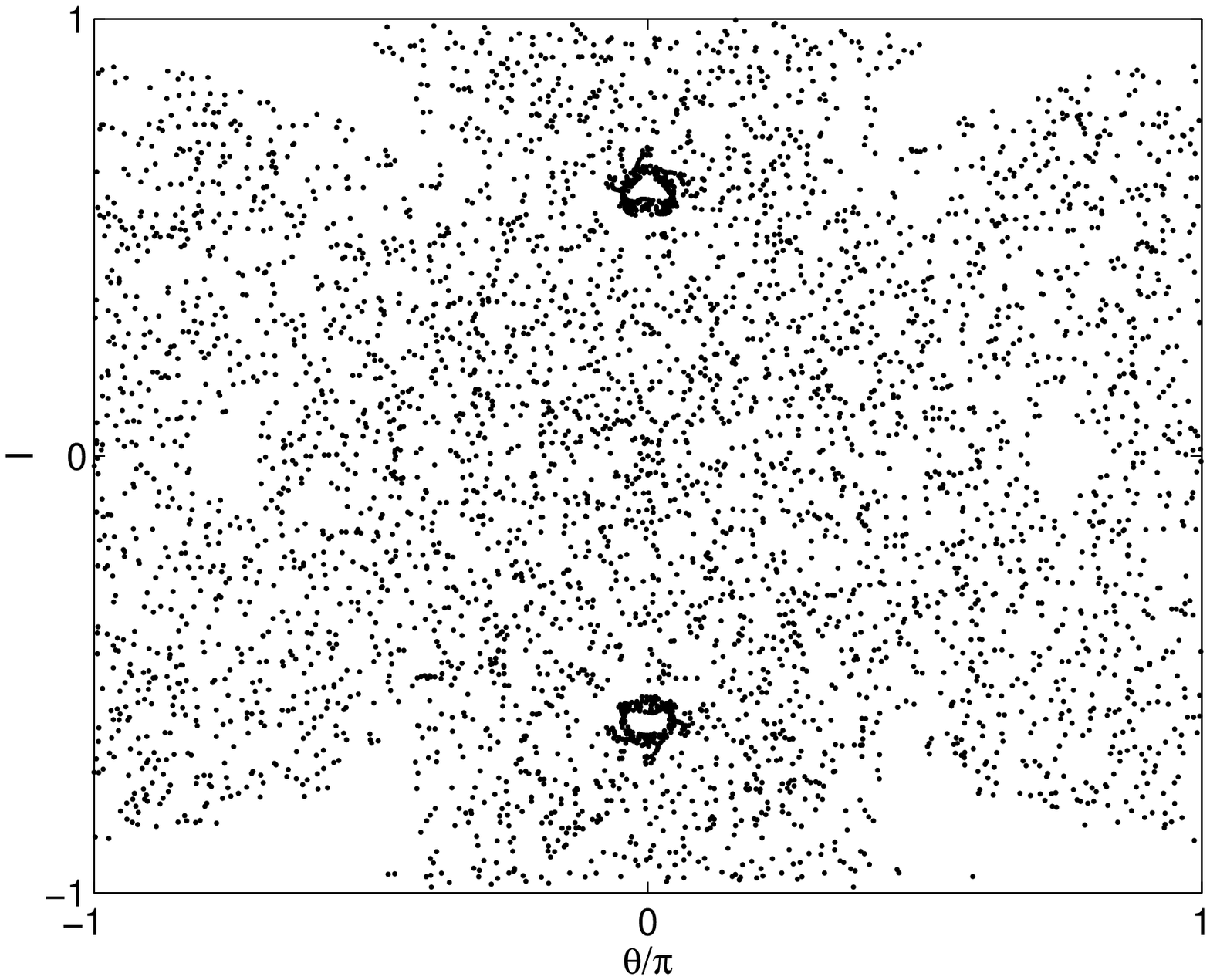}
\includegraphics[width=4.0cm, clip]{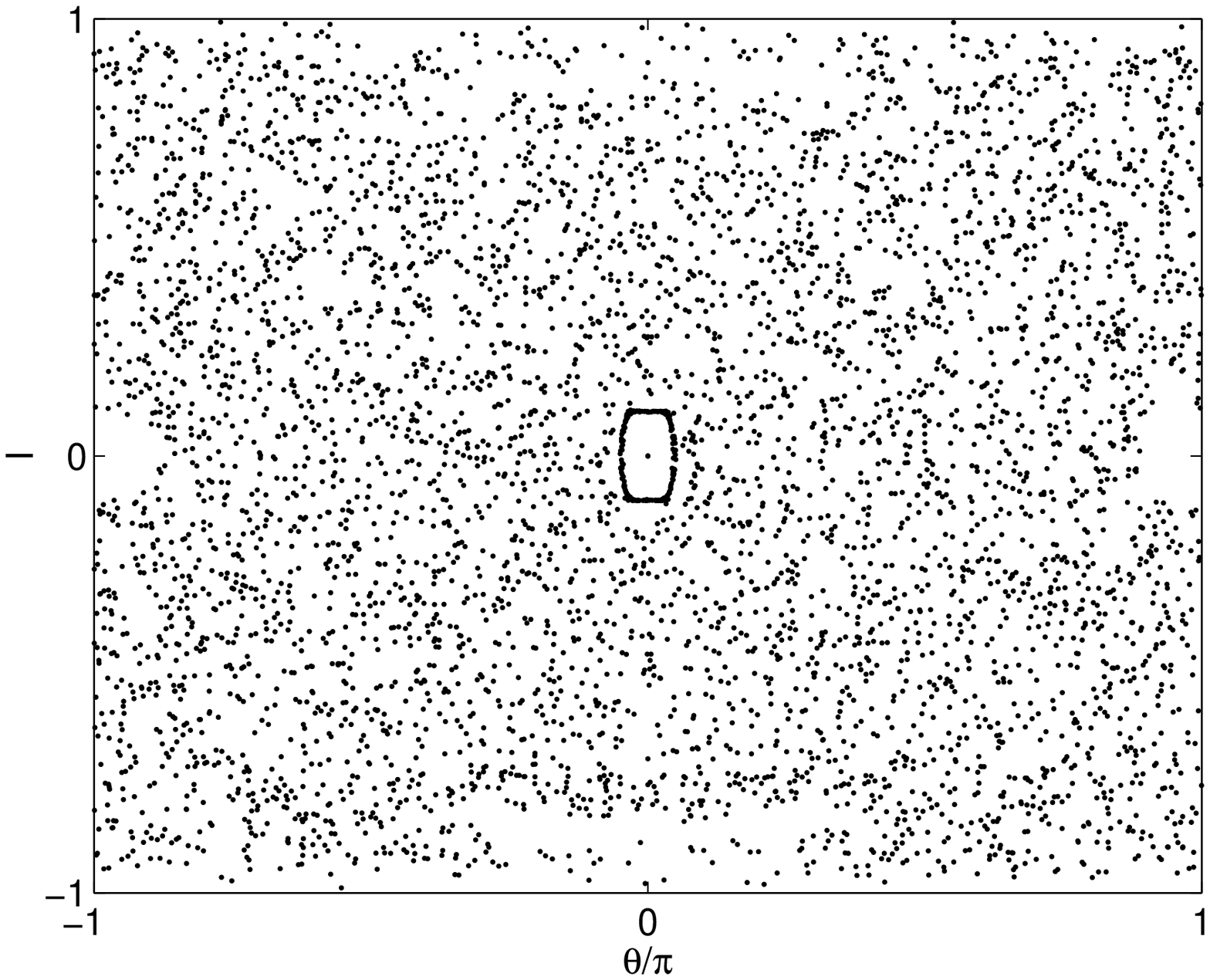}
\caption{Phase portrait (stroboscopic map) of the system (\ref{7})
for $g=0.4$ and $F=2.0$ (upper left panel), $F=1.3$ (upper right),
$F=0.7$ (lower left), and $F=0.5$ (lower right panel).}
\label{fig4}
\end{figure}

One obtains a useful  insight in physics of the discussed phenomenon
by considering the limiting case of the lattice with only two sites,
\begin{equation}
\label{6}
H(t)=-J\cos(\omega_B t)(a^*_2a_1+a^*_1a_2)
    +\frac{g}{2}\sum_{l=1}^2 |a_l|^4 \;,
\end{equation}
where $|a_1|^2+|a_2|^2=2$ is the integral of the motion.
Using the substitution $a_{1,2}=\sqrt{2I_{1,2}}\exp(i\theta_{1,2})$,
$I=I_1-I_2$, $\theta=\theta_1-\theta_2$, one maps (\ref{6})
to the periodically driven pendulum,
\begin{equation}
\label{7}
H(t)=gI^2-J\sqrt{1-I^2}\cos(\omega_B t)\cos\theta \;,
\end{equation}
with the angular momentum $I$ resticted to the interval
$|I|\le1$ \cite{remark}. The stroboscopic map of the pendulum (\ref{7})
is depicted in Fig.~\ref{fig4} for different values of the static force
$F=2.0$, $1.3$, $0.7$, $0.5$. The periodic solution (\ref{3}),
which is of our particular interest, corresponds in Fig.~\ref{fig4} 
to the stationary point $(I,\theta)=(0,0)$. As $F$ is decreased,
the stability island surrounding this elliptic point shrinks to zero
and at $F\approx0.9$ (see inset in Fig.~\ref{fig1}) it
bifurcates into the hyperbolic point.  We also note
that for $F<0.9$ the phase space of the system is dominated by
the chaotic component. In this sense the onset of dynamical
instability at $F\approx0.9$ corresponds to the condition of developed 
chaos \cite{jetp5}.

\begin{figure}[t!]
\center
\includegraphics[width=8.5cm, clip]{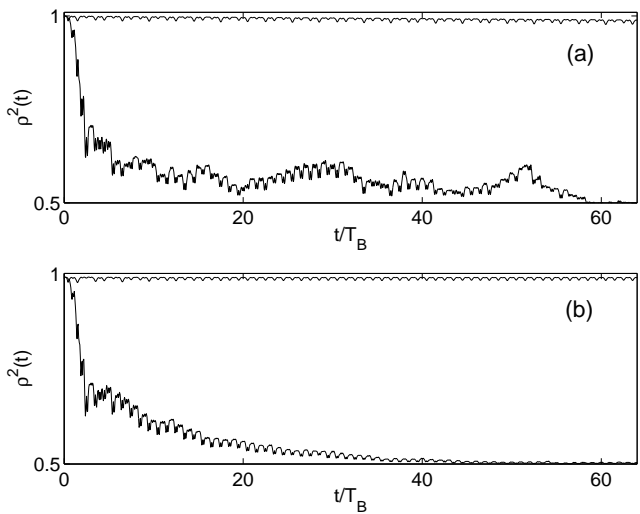}
\caption{Upper panel: Dynamics of $\rho^2(t)={\rm Tr}[\hat{\rho}^2(t)]$
for the system (\ref{8}) with ${\bar n}=50$ and $g={\bar n}W=0.4$.
Two curves corresponds to $F=1.3$ [$\rho^2(t)\approx1$] and
$F=0.7$ [decay of $\rho^2(t)$]. The lower panel is the `classical'
simulation of the decoherence process.}
\label{fig5}
\end{figure}

Next we discuss the relation between the onset of chaos and
decoherence (depletion) of BEC. To address this problem
we need to treat the system microscopically, where for
the discussed two-site model the microscopic Hamiltonian
is given by the time-dependent Bose-Hubbard model,
\begin{equation}
\label{8}
\widehat{H}(t)=-J\cos(\omega_B t)(\hat{a}^\dag_2\hat{a}_1+h.c.)
+\frac{W}{2}\sum_{l=1}^2\hat{n}_l(\hat{n}_l-1) \;.
\end{equation}
As an overall characteristic of decoherence we consider
the trace of the squared density matrix, $\rho^2(t)
={\rm Tr}[\hat{\rho}^2(t)]$, where
\begin{equation}
\label{10}
\hat{\rho}_{l,m}(t)=N^{-1}\langle\Psi(t)|\hat{a}^\dag_{l}
\hat{a}_{m}|\Psi(t)\rangle
\end{equation}
is the one-particle density matrix. (We recall that for a
coherent evolution $\rho^2(t)=1$.)
In Eq.~(\ref{10}) $|\Psi(t)\rangle$ is the total $N$-particle
wave function of the system ($N=100-400$ in the numerical simulations)
and as an initial condition $|\Psi(0)\rangle$ we choose
the ground state of (\ref{8}) for $F=0$, which is approximately given
by the product of $N$ Bloch waves with zero qusimomentum.
It is found that when the central stability island in Fig.~\ref{fig4}
is large enough in comparison with the quantum of the phase
space volume $2\pi\tilde{\hbar}$ \cite{remark0} decoherence is negligible.
On the contrary, when the periodic point $(I,\theta)=(0,0)$ is hyperbolic, 
we observe rapid decoherence of the system (see upper panel
in Fig.~\ref{fig5}). Note that the rate of decoherence can be well 
estimated by using the classical (i.e., mean field)
dynamics and considering an ensemble of initial conditions
scattered around the periodic point. Then the one-partical density
matrix (\ref{10}) is given by the correlation matrix,
\begin{equation}
\label{11}
\rho_{l,m}(t)=L^{-1}\langle\langle a^*_l(t)a_m(t)\rangle\rangle \;,
\end{equation}
where the double-angle brackets denote the average over an ensemble
of the initial conditions. The characteristic width of the
distribution function for the action and angle variables is obviously
given by the effective Planck's constant or, equivalently, by the
quantum fluctuations of the number of atoms in one well,
$\overline{\Delta I^2}\sim \tilde{\hbar}^2\overline{\Delta n^2}\sim1/{\bar n}$,
and phase fluctuations,
$\overline{\Delta \theta^2}\sim1/\overline{\Delta n^2}\sim1/{\bar n}$
\cite{remark1}. For ${\bar n}=50$ the classical dynamics of $\rho^2(t)$
is depicted in the low panel of Fig.~\ref{fig5}. By comparing both
panels one concludes that decoherence of the quantum system (\ref{8})
is actually due to the chaotic dynamics of its classical counterpart (\ref{6}).

\section{Bloch oscillations}
\begin{figure}[t!]
\center
\includegraphics[width=8.5cm, clip]{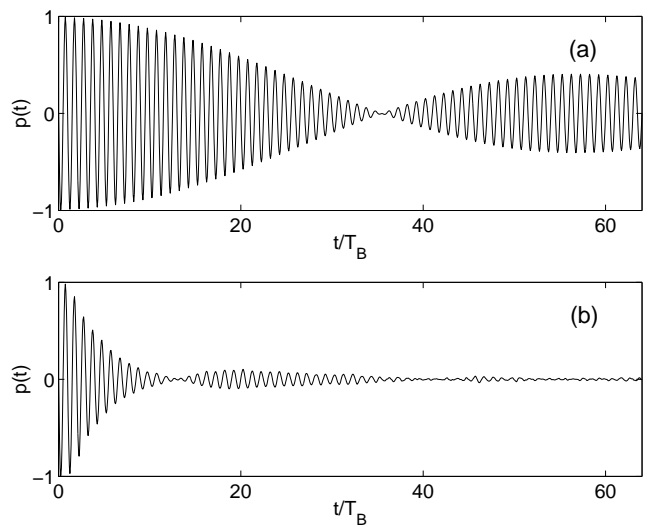}
\caption{Bloch oscillations of the mean atomic momentum
for $F=1.3>F_{cr}$ (a) and $F=0.7<F_{cr}$ (b). [Average over
100 different realizations of the initial conditions (\ref{12}) with
${\bar n}=200$.] The width of the atomic array is $\sigma=26.5$,
the nonlinear parameter $g={\bar n}W=0.4$.}
\label{fig6}
\end{figure}

Although it looks problematic to extend the above phase-space analysis 
of Sec.~\ref{sec3} for $L\gg1$, 
some conjectures are still possible. Indeed, using the polar
presentation for complex amplitudes $a_l$, one maps the $L$-site system into
a system of $L$ driven coupled pendula and there is no doubt that
this system would be generally chaotic \cite{Thom03}. On the other hand,
absence of the dynamical instability for $F>F_{cr}$ indicates the presence
of the stable periodic trajectory in the multi-dimensional phase
space of the system corresponding to the stable BO. The crucial point is,
however, whether the stability island surrounding this periodic trajectory
is large enough to support the quantum state. To answer this question
we numerically solve the DNLSE for randomly varied initial conditions,
\begin{displaymath}
|a_l|^2=\overline{|a_l|^2}+\xi\left(\frac{\overline{|a_l|^2}}
{\bar n}\right)^{1/2} \;,\quad
\overline{|a_l|^2}=\exp\left(-\frac{l^2}{\sigma^2}\right) \;,
\end{displaymath}
\begin{equation}
\label{12}
\theta_l=\xi \left(4{\bar n}\overline{|a_l|^2}\right)^{-1/2} \;,\quad
a_l=|a_l|\exp(i\theta_l)
\end{equation}
where $\xi$ is a random Gaussian variable with $\overline{\xi^2}=1$,
which accounts for the quantum fluctuations. The results are presented
in Fig.~\ref{fig6}, which shows dynamics of the mean atomic momentum
for $g=0.4$, ${\bar n}=200$, $\sigma=25.6$,
$F=1.3$ (upper panel) and $F=0.7$ (lower panel).
It is seen that the stable regime of BO is also stable against
the quantum fluctuations. This validates our conclusion about two
qualitatively different regimes of BO.

It interesting to estimate critical values of
the parameters for a typical laboratory experiment.
Taking as an example the recent experiment \cite{Roat04}
with $^{87}$Rb atoms in a vertically oriented quasi one-dimensional
lattice of the depth 4 recoil energies, one has $dF=0.28$, $J=0.17$,
and $W=0.71\cdot10^{-5}$ recoil energies, respectively. Then the condition
on the critical magnitude of the static force is formulated as a condition
on the critical density of the condensate and corresponds to
${\bar n}_{cr}=120$ atoms per lattice site. Since the peak density in
Ref.~\cite{Roat04} is larger than ${\bar n}_{cr}$, BOs should
decay, which agrees with the experimental finding.

\section{Conclusion}
We have studied the dynamics induced by a static force of a BEC 
of cold atoms in an optical lattice. Depending on the static force
magnitude $F$ (or, equivalently, on the density of the condensate ${\bar n}$)
the system is shown to have two qualitatively different regimes of 
BO -- exponential decay of oscillations for
$F<F_{cr}({\bar n})$ and `quasiperiodic' oscillations
for $F>F_{cr}({\bar n})$. It is argued in the paper that the former
regime reflects the chaotic dynamics of the system, where the
transition to chaos coincides with onset of the dynamical
instability in DNLSE.

It is worth of noting that the above results are obtained within the
mean field approach. Because this approach assumes the limit
${\bar n}\rightarrow\infty$, $W{\bar n}=g/{\bar n}\rightarrow0$,
for any finite ${\bar n}$ the classical (macroscopic) dynamics of the system,
governed by DNLSE, deviates from the quantum (microscopic) dynamics, governed 
by the Bose-Hubbard model. It is a challenge both for theory and
experiment to establish the relation between the classical and quantum 
results, depending on the parameter $\tilde{\hbar}=1/{\bar n}$.
We would also like to stress that for considering the problem
of quantum-to-classical correspondence one has to average the solution
of DNLSE over an appropriate ensemble of the initial conditions.
Without this additional procedure, the solutions of the DNLSE may have
nothing to do with the solutions of the Bose-Hubbard model.

In this present paper we analyse the problem of quantum-to-classical
correspondence for BO by using a two-site model. Both the similarities
and discrepancies were found. In particular, we show that the classically
chaotic dynamics of the system for $F<F_{cr}$ is responsible for 
decoherence of the system, which is the deep reason for the decay of BOs.
At the same time, the rate of decoherence, obtained by the means of the DNLSE,
appears to be correct only for the short-time dynamics.

To conclude, we briefly discuss the relation of the above results
with those of our recent papers \cite{prl57,pre60,prl61}, devoted
to the Bloch dynamics of the system in the deep quantum regime 
$\tilde{\hbar}\sim1/{\bar n}\sim1$ \cite{remark2}. Clearly, in this case
one cannot appeal to the results of the classical (DNLSE) analysis
and the system should be treated microscopically. Remarkably, 
that this microscopical analysis also reveals two regimes of BO 
-- regular (not-decaying) oscillations for a strong forcing \cite{prl57}
and rapid decay (decoherence) of BO in the case of a weak static 
force \cite{prl61}.

The author acknowledges discussions with J.~Brand, A.~Buchleitner, S.~Flach, 
and S.~Sinha.



\begin{thebibliography}{10}
\bibitem{remark0}
The effective Planck's constant $\tilde{\hbar}$ is given by 
$\tilde{\hbar}=2/{\cal N}$, where ${\cal N}=2N+1\approx4{\bar n}$
is dimension of the Hilbert space.
\bibitem{remark1}
More regorously, the distribution function for the classical ensamble
is given by Husimi image of the initial quantum state.
\bibitem{remark2}
This regime is typically realised in 3D lattices. Note that the microscopic
interaction constant $W$ in 3D lattices is typically three
order of magnitude larger than in the quasi 1D lattices which,
together with ${\bar n}\sim1$, gives approximately the
same value for the nonlinear parameter $g$.
\end{thebibliography}
\end{document}